\documentstyle[11pt,cs11,html,epsf]{article}

\begin{document}

\title{(Teff,log g,[Fe/H]) Classification of Low-Resolution Stellar
Spectra using Artificial Neural Networks}

\author{Shawn Snider, Yuan Qu, Carlos Allende Prieto}
\affil{McDonald Observatory and Department of Astronomy, The University of Texas  at Austin RLM 15.308, Austin, TX 78712-1083,  USA}

\author{Ted von Hippel}
\affil{Gemini Observatory, USA}

\author{Timothy C. Beers}
\affil{Department of Physics and Astronomy, Michigan State University,  East Lansing, MI 48824, USA}

\author{Chistopher Sneden, David L. Lambert}
\affil{McDonald Observatory and Department of Astronomy, The University of Texas  at Austin RLM 15.308, Austin, TX 78712-1083,  USA}

\author{Silvia Rossi}
\affil{Instituto Astronomico e Geofisico, Universidade de Sao Paulo, 
Av. Miguel Stefano 4200, 04301-904, Sao Paulo, SP, Brazil}

\begin{abstract}

New generation large-aperture telescopes, multi-object spectrographs,
and large format detectors are making it possible to acquire very large
samples of stellar spectra rapidly.  In this
context, traditional star-by-star spectroscopic analysis are no longer
practical.~ New tools are required that are capable of extracting quickly and
with reasonable accuracy important basic stellar parameters coded in the
spectra.~ Recent analyses of Artificial Neural Networks (ANNs) applied to the
classification of~ astronomical spectra have demonstrated the ability of this
concept to derive estimates of~ temperature and luminosity. We have adapted the back-propagation ANN technique
developed by von Hippel et al. (1994) to predict effective temperatures,
gravities and overall metallicities from~ spectra with resolving power
$\lambda/\delta\lambda \simeq 2000$ and low signal-to-noise ratio. We show that ANN techniques
are very effective in executing a three-parameter (Teff,log g,[Fe/H]) stellar
classification. The preliminary results show that the technique is even
capable of identifying outliers from the training sample.
\end{abstract}


\section{Introduction}

Artificial Neural Networks (ANNs) have been~ applied to the
classification of stellar spectra very recently, but with great
success. These computational systems provide a mapping from a set of
inputs to a set of desired outputs and can be~ trained to classify~
anything with great accuracy and speed. Vieira \& Ponz (1995) made use
of this technique to carry out the spectral classification~ of
low-resolution spectra obtained by the IUE satellite. They found that
ANNs performed better than classical methods based on a defined metric
distance, making it possible an accuracy of 1.1 spectral subclasses.
More recently, Bailer-Jones et al. (1998) trained an ANN to classify
objective-prism spectra from the Michigan Spectral Survey, extracting
the spectral type (std. deviation = 1.09) and the luminosity class
(success rate $>$ 95\%).

We have stepped forward from the two-dimensional (temperature and
luminosity class) classification to the three-dimensional, including
the stellar metal content. We have made use of part of the
observational material collected by Beers and his large collaborative
projects (Beers et al. 1999). Table~\ref{table1} summarizes the main
characteristics of the spectra and the acquisition places.

\begin{table}\caption{Origen and description of the spectra employed in the study} \label{table1}
\begin{center}\scriptsize
\begin{tabular}{lllll}
\multicolumn{5}{c}{\bf ~ } \\
\multicolumn{1}{c }{ } & \multicolumn{1}{c }{Telescope } & \multicolumn{1}{c
}{Detector }  & \multicolumn{1}{c }{Coverage ({\AA}) } & \multicolumn{1}{c }{{\AA}/pix} \\
\tableline
{ } & {Mount Stromlo Observatory 1.9 m } & {PCA } &  \multicolumn{1}{c }{3750-4100 } & {0.4  } \\
{ } & {Siding Spring Observatory 2.3 m } & {PCA } &  \multicolumn{1}{c }{3800-4300 } & {0.5  } \\
{ } & {Siding Spring Observatory 2.3 m } & {Loral 1024$\times$1024 } & \multicolumn{1}{c }{3800-4400 } & {0.5  } \\
{ } & {Siding Spring Observatory 2.3 m } & {SITe 1752$\times$532 } &
 \multicolumn{1}{c }{3750-4600 } & {0.5  } \\
{ } & {Las Campanas 2.5 m } & {Reticon }  &
\multicolumn{1}{c }{3700-4500 } & {0.3 (0.65)  } \\
{ } & {2D-FRUTTI } & {0.6 (0.65)  } & &  { } \\
{ } & {Palomar 5 m } & {Reticon }  &
\multicolumn{1}{c }{3700-4500 } & {0.3 (0.65)  } \\
{ } & {2D-FRUTTI } & {0.6 (0.65)  } &  &  { } \\
{ } & {European Southern Observatory 1.5 m } & {Ford 2048$\times$2048 } &
 \multicolumn{1}{c }{3750-4750 } & {0.65  } \\
{ } & {European Southern Observatory 1.5 m } & {Loral 2048$\times$2048 }  & \multicolumn{1}{c }{3750-4600 } & {0.5  } \\
{ } & {Kitt Peak National Observatory 2.1 m } & {Tek 2048$\times$2048 } &
 \multicolumn{1}{c }{3750-5000 } & {0.65  } \\
{ } & {Isaac Newton 2.5 m } & {Tek 1024$\times$1024 }  & \multicolumn{1}{c }{3750-4700 } & {0.9  } \\
{ } & {Lowell Observatory 1.8 m } & {Tek 512$\times$512 }  & \multicolumn{1}{c }{4000-4250 } & {1.0  } \\
{ } & {Observatoire Haute-Provence 1.9m } & {Tek 512$\times$512 } &
\multicolumn{1}{c }{3750-4250 } & {0.9 }
\end{tabular}
\end{center}
\end{table}


\section{Input data}

\begin{figure}
\plotfiddle{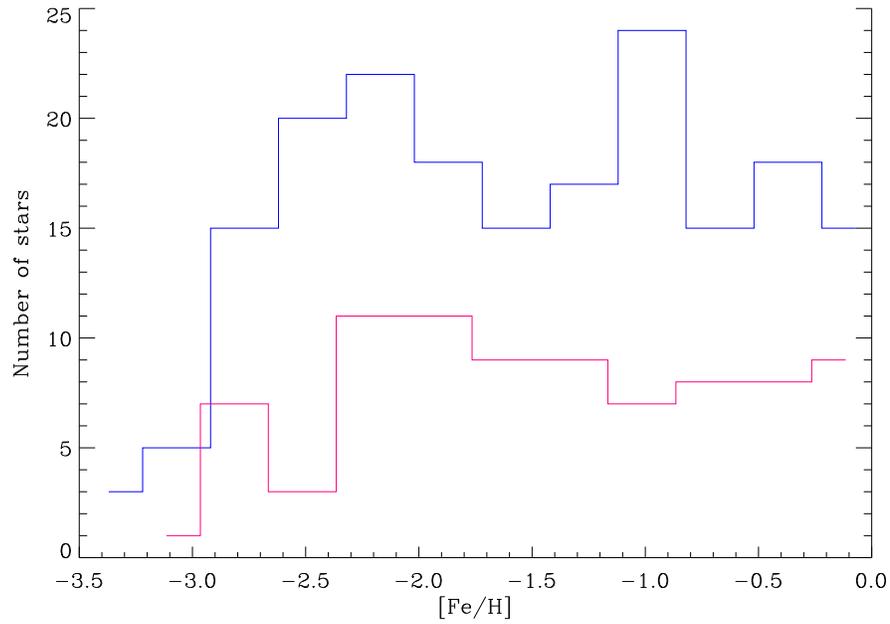}{7cm}{90}{50}{50}{200}{-30}
\caption{Distribution of metallicities in the {\it training} (blue) and {\it testing} (red) samples} \label{fig1}
\end{figure}

A selection of 182 stars spanning all metallicites, gravities and effective
temperatures (Teffs) was selected for training. ANNs can over-learn, that is,
they may get to the level of taking into account features that are particular
to the stars in the training sample, rather than typical characteristics of
the spectral classes, gravities, and metallicities they represent. For this
reason, an independent sample must be used to check that the net is properly
classifying the stars. 82 stars were used for this purpose. Figure~\ref{fig1}
 shows the distribution of the metallicity of the training and testing
samples. 

\begin{figure}
\plotfiddle{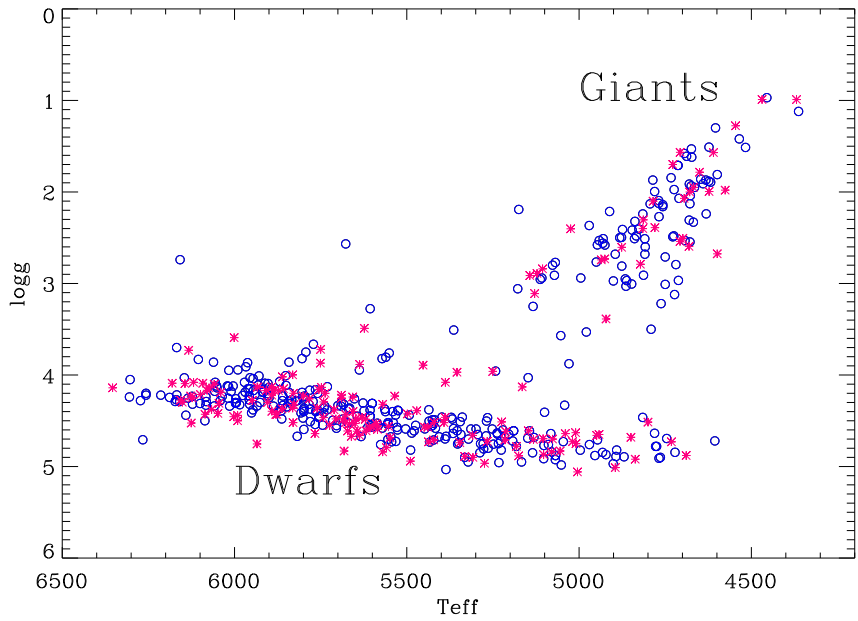}{7cm}{90}{50}{50}{200}{-30}
\caption{Distribution of the {\it training} (blue) and {\it testing} (red) samples across the HR diagram} \label{fig2}
\end{figure}

The training and testing samples were selected to make sure that the mapping
of the Teff-logg was adequate as well, as demonstrates Figure~\ref{fig2}. 

Metallicities for the testing and training samples were compiled by Beers et
al. (1999). Effective temperatures have been derived~ from compiled B-V
colors, applying the calibrations of Alonso et al. (1996) for dwarfs and
subgiants, and Alonso et al. (1999, private communication) for more evolved
stars. These calibrations are based on the InfraRed Flux Method, developed by
Blackwell and collaborators (e.g., Blackwell \& Lynas-Gray 1994). We have
taken advantage of the distance estimates made by~ Beers et al. (1999)
(spectroscopic parallaxes) to interpolate in the evolutionary isochrones of
Bergbush \& VandenBerg (1992) and derive bolometric corrections and masses.
The calculated luminosities were combined with the effective temperatures to
obtain the stellar radii, and then with the masses to estimate the gravities. 

\section{Spectra processing}

Before entering the ANN, the spectra were pre-processed using IRAF routines.
They were first continuum flattened, using a 3rd order spline interpolation
method. Then, the spectra were shifted to a pre-chosen template velocity by
the ``Fxcor'' and ``Dopcor'' packages and rebinned to a common dispersion
(0.646 {\AA}/pix). Finally, using ``Wspectext'' , the spectra were converted
into text format.

\section{Applying the neural network}

All weights are initially random. A node fires at a value given by a sigmoid
function, {\bf F(a) = 1/(1+e\raisebox{.6ex}{-a})}, where {\bf a = $\Sigma$
(w\raisebox{-.6ex}{ij} x I\raisebox{-.6ex}{i})},{\bf }where {\bf
w\raisebox{-.6ex}{ij }}is the corresponding weight and {\bf I\raisebox{-.6ex}{i
}}the corresponding input. Then {\bf F(a) = O\raisebox{-.6ex}{j}}, the hidden
(or output) node value. 

The weight training is accomplished by means of the Ripley code (Ripley
1993), a quasi-Newtonian optimization method.~ Besides the initial
random weights, Ripley's code eliminates all free parameters that are
present in most back propagation networks (e.g. learning rate, momentum
term).

Artificial Neural Networks of 3, 5, 7, and sometimes 9 and 11 hidden
nodes were tried with varying random weight initializations. The net
architecture finally~ used is 1 hidden layer and 5 hidden nodes, which
produced the most reasonable results based on overdetermination as well
as reliability and time constraints.~ A typical training session
involved about 1000 iterations in perhaps 30 minutes on a Sun ultra 30.
This implies a testing time of much less than 1 second per spectrum.

We chose a final spectral range of 3630 to 4890 {\AA} before running the ANN
to ensure the best spectral quality possible. At 0.646 {\AA}/pix, this yields
1952 spectral resolution elements, i.e. input values, per spectrum. 

\section{Results and conclusions}

\begin{figure}
\plotfiddle{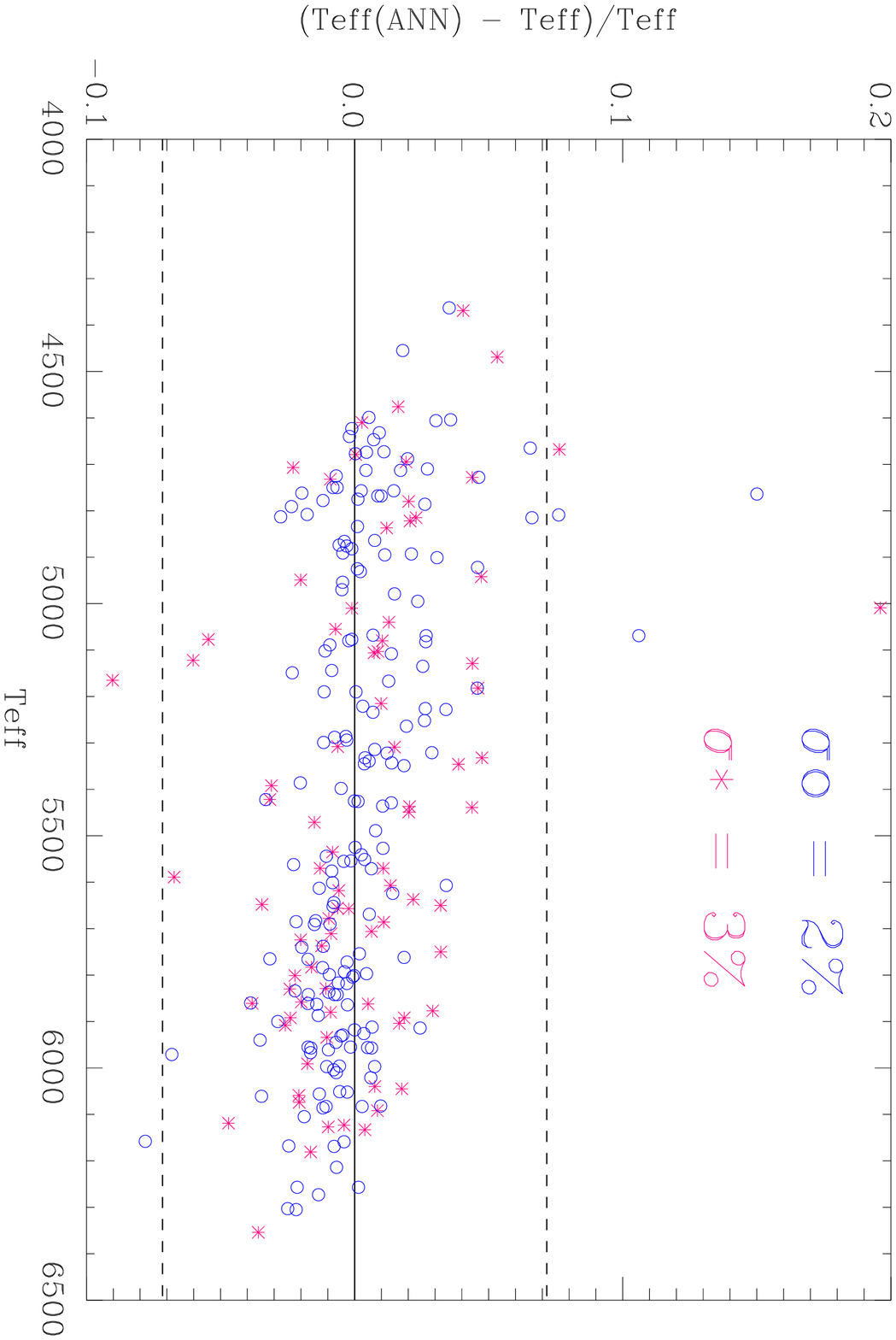}{5.5cm}{90}{40}{35}{150}{0}
\plotfiddle{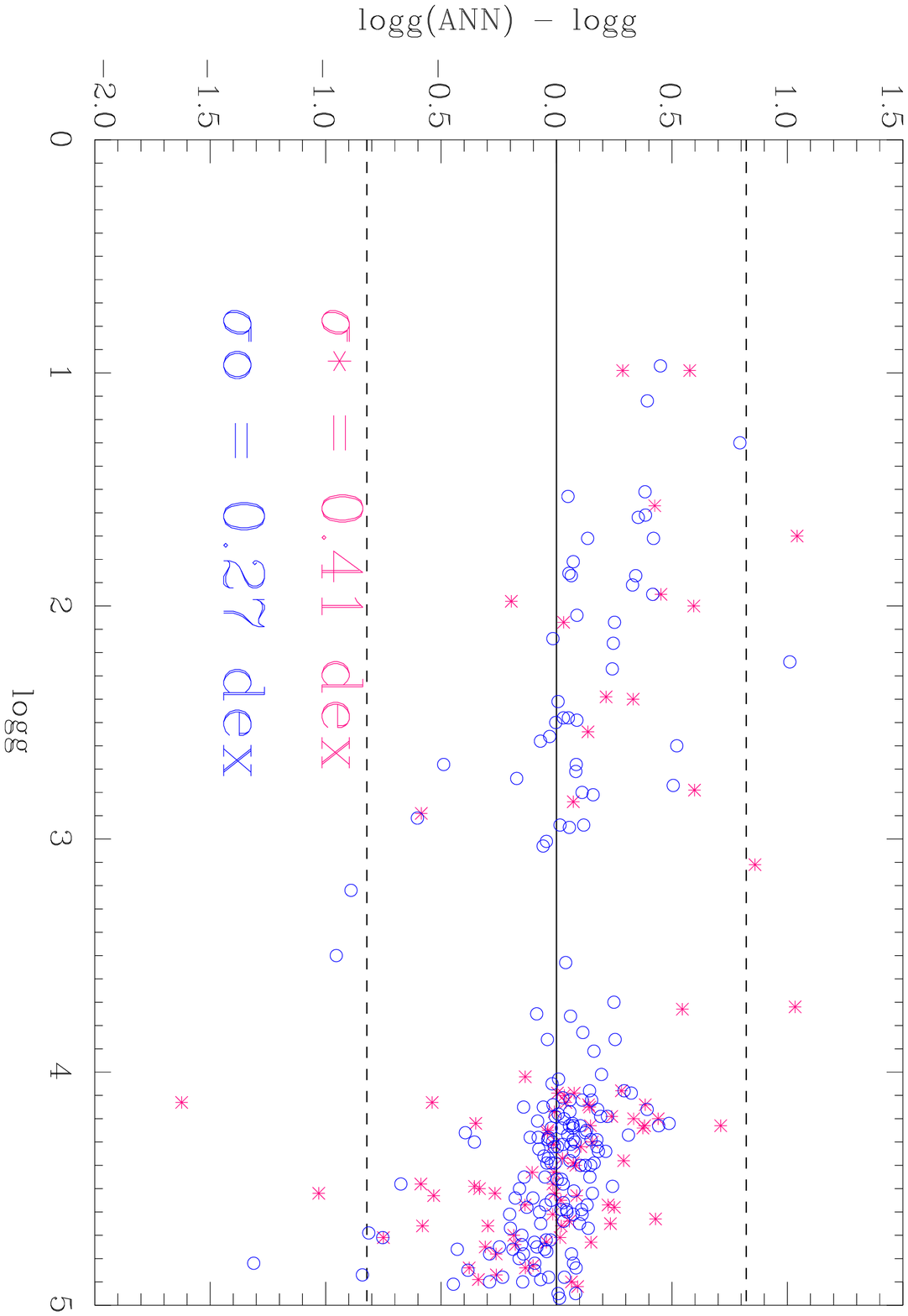}{5.5cm}{90}{40}{35}{150}{0}
\plotfiddle{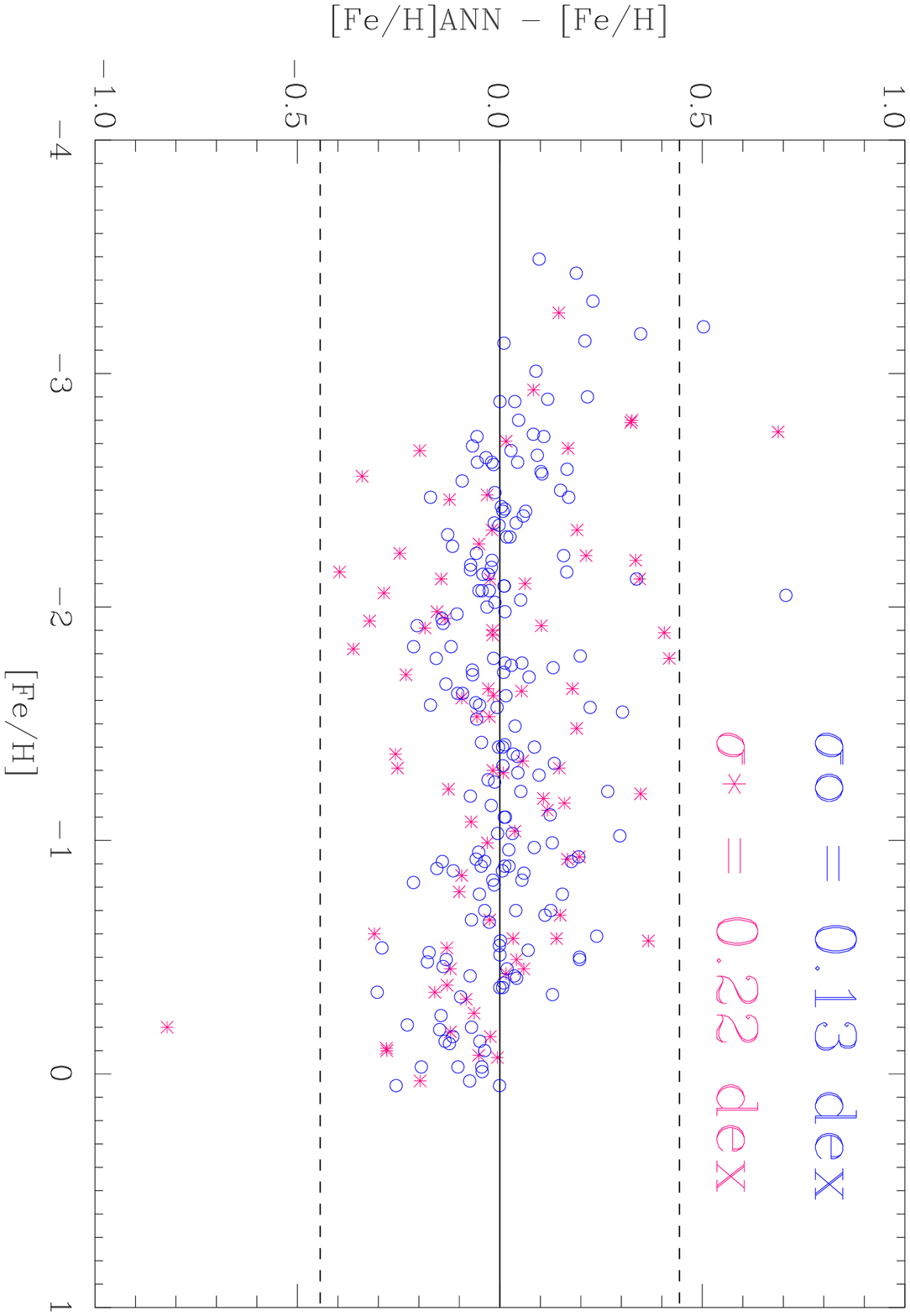}{5.5cm}{90}{40}{35}{150}{0}
\caption{Deviation of the ANN answers from the assumed parameters for the {\it training} (blue) and {\it testing} (red) samples} \label{fig3}
\end{figure}

Our results are displayed in Figure~\ref{fig3}. Several stars in the training sample were indicated by the net as problematic.
A close look to those outliers revealed that an important number of them
corresponded to obvious errors in the application of the spectroscopic
parallax technique. They have been excluded from the comparison shown here,
and will be included, with the corrected stellar parameters,~ in future ANN
training runs. Other outliers for which no obvious explanation was found, have
been kept in the comparison.

The performance of the trained ANNs can be graphically seen in the following
graphs, and is summarized in the Table~\ref{table2}, where the rms differences
 between the known parameters and those provided by the net~ are
displayed. The information for the training sample provides a glimpse on how
well the ANN is learning.

\begin{table}\caption{Rms differences between the assumed stellar parameters and those provided by the ANN for the {\it training} and {\it testing} samples}\label{table2}
\begin{center}
\begin{tabular}{lll}
{Parameter } & {$\sigma$(Training) } & {$\sigma$(Testing)  } \\
\tableline
{Teff } & {125 K } & {186 K  } \\
{logg } & {0.27 dex } & {0.41 dex  } \\
{[Fe/H] } & {0.13 dex } & {0.22 dex }
\end{tabular}
\end{center}
\end{table}

The few outliers mainly represent unusual spectra which are either: 

~~ a) under-represented by the training set, or 

~~ b) have poor quality spectra and/or have been unreasonably continuum
flattened.

We expect future ANN runs to provide better results, after correcting~ errors
that have been already identified, and others yet to be investigated,~ in the
parameters adopted for at least some of the outliers.

\acknowledgments  This work has been partially funded by  the U.S. NSF (grant AST961814), and the Robert A. Welch Foundation of Houston, Texas.


\begin{references}
\reference Alonso, A., Arribas, S., \& Mart\'{\i}nez Roger, C. 1996, \aap, 313, 873  
\reference Bailer-Jones, C. A. L. 1997, \pasp, 109, 932 (PhD thesis abstract)
\reference Bailer-Jones, C. A. L., Irwin, M., \& von Hippel, T. 1998, \mnras, 298, 361
\reference Beers, T. C., Rossi, S., Norris, J. E., Ryan, S. G., \& Shefler, T. 1999, \aj, 117, 981
\reference Bergbusch, P. A., \& Vandenberg, D. A. 1992, \apjs, 81, 163
\reference Blackwell, D. E., \& Lynas-Gray, A. E.   1994, \aap, 282, 899
\reference Vieira, E. F., \& Ponz, J. D. 1995, \aaps, 111, 393
\reference von Hippel, T., Storrie-Lombardi, L. J., Storrie-Lombardi, M. C., \& Irwin, M. J. 1994, \mnras, 269, 97


\end{references}
\end{document}